\documentclass[10pt,a4paper]{article}
\usepackage[T1]{fontenc}
\usepackage{amsmath,amsfonts, amssymb, amsbsy, mathrsfs, mathtools, bbm, enumitem}
\usepackage[medium]{titlesec}
\usepackage{float}
\usepackage{makeidx}
\usepackage{graphicx}
\usepackage{color}
\usepackage{url}
\usepackage{authblk}
\usepackage{multirow}
\usepackage{lscape}
\usepackage{tensor}
\usepackage[font=small,labelfont=bf]{caption}

\usepackage{mathpazo}
\usepackage[toc]{appendix}
\usepackage[hidelinks,colorlinks=true,linkcolor=red,citecolor=blue,urlcolor=magenta]{hyperref}
\usepackage[left=2.5cm, right=2.5cm, top=3.0cm, bottom=3.0cm]{geometry}

\usepackage{fancyhdr}

\newcommand{\email}[1]{\href{mailto:#1}{#1}}

\newcommand{\df}{\textrm{d}}

\newcommand{\cg}{\textnormal{\textsl{g}}}

\newcommand{\hf}{{\frac{1}{2}}}

\newcommand{\la}{\left\langle}
\newcommand{\ra}{\right\rangle}

\renewcommand{\bar}{\overline}

\titleformat{\section}
{\normalfont\large\bfseries}{\thesection}{1em}{}

\titleformat{\subsection}
{\normalfont\normalsize\bfseries}{\thesubsection}{1em}{}

\titleformat{\paragraph}
{\normalfont\normalsize\bfseries}{\theparagraph}{1em}{}

\titleformat{\subparagraph}
{\normalfont\normalsize\bfseries}{\thesubparagraph}{1em}{}

\usepackage[square,numbers,sort&compress]{natbib}

\numberwithin{equation}{section}

\allowdisplaybreaks[1]

\begin{document}
	\setlength{\bibsep}{0pt}
	
	\title{ \textbf{\Large On the Effects of Non-metricity in an Averaged Universe}}
	\author[1,2]{\normalsize Anish Agashe\thanks{\email{anagashe@smcm.edu}}}
	\author[2]{\normalsize Sai Madhav Modumudi\thanks{\email{saimadhav.modumudi@utdallas.edu}}}
	\affil[1]{{\small \textit{Department of Physics, St. Mary's College of Maryland,}
			
			\textit{47645 College Dr., St. Mary's City, Maryland, USA 20686}}}
	\affil[2]{ {\small \textit{Department of Physics, University of Texas at Dallas,}
			
			\textit{800 W Campbell Rd., Richardson, Texas, USA 75080}}}
	\date{}

	\maketitle
	
	\begin{abstract}
		In the covariant averaging scheme of macroscopic gravity, the process of averaging breaks the metricity of geometry. We reinterpret the back-reaction within macroscopic gravity in terms of the non-metricity of averaged geometry. This interpretation extends the effect of back-reaction beyond mere dynamics to kinematics of geodesic bundles. With a 1+3 decomposition of the spacetime, we analyse how geometric flows are modified by deriving the Raychaudhuri and Sachs equations. We also present the modified forms of Gauss and Codazzi equations. Finally, we derive an expression for the angular diameter distance in Friedmann Lema\^itre Robertson Walker universe and show that non-metricity modifies it only through the Hubble parameter. Thus, we caution against overestimating the influence of back-reaction on the distances.\\
		
		\noindent {\small \textit{Keywords}: Non-metricity; Back-reaction; Averaging problem; 1+3 formalism}
	\end{abstract}

	{
	}
	\section{Introduction}
	
	The universe is traditionally modelled by employing cosmological solutions to the Einstein field equations. In these models, the matter distribution is taken to be averaged over cosmological scales, and hence the Einstein tensor needs to be averaged as well \cite{ellis2,ellis3}. However, averaging tensor fields on a differentiable manifold is not straightforward and is a well-known problem in relativistic cosmology \cite{ellisbr,hoogen1}. The difficulty arises because of the lack of a mathematically rigorous integration operation on non-Euclidean geometries. Moreover, the non-linearity of the Einstein field equations exacerbates the problem since the averaged dynamics of a given metric are different from the dynamics of an averaged metric, i.e., $ \la \boldsymbol{G}[\textsl{\textbf{g}}] \ra \ne \boldsymbol{G}[\la\textsl{\textbf{g}}\ra] $. This is referred to as the averaging problem in the literature. There is a long history of approaches towards this problem (see \cite{ellisbr,cliftonbr,buchertbr} and the references within). In all the approaches, however, the problem is effectively addressed by modifying the dynamics of the averaged spacetime. This modification may take different forms in various approaches, and is generically referred to as cosmological back-reaction \cite{cliftonbr,buchertbr}. One such approach is macroscopic gravity \cite{zala1,zala2,zala3,zala4} which offers a non-perturbative and fully covariant solution to the averaging problem.
	
	Macroscopic gravity calculates space-time averages of arbitrary tensor fields through averaging bi-vectors and Lie dragging of the averaging regions. When this averaging scheme is applied to general relativity (GR), the Riemannian\footnote{The signature is implicitly assumed to be Lorentzian. Therefore, the term `Riemannian' here is used to mean \textit{pseudo-}Riemannian.} structure of the geometry is broken. The averaged geometry (the geometry at the macroscopic scales) is equipped with a metric-incompatible but symmetric connection that arises from averaging. This leads to a non-vanishing non-metricity but a vanishing torsion. The dynamics of macroscopic gravity is given by taking a space-time average of the Einstein field equations. The back-reaction in these dynamics is also given in terms of the non-metricity, as has been noted in \cite{zala6,hoogen3}.
	
	The aim of this work is to analyse the effects of non-metricity in macroscopic gravity on various aspects of relativistic cosmology. We do this by treating macroscopic gravity as a metric-affine theory. In such a theory, the geometry is endowed with a metric and an independent connection which is precisely how macroscopic gravity proposes the averaged geometry to be. This means that the back-reaction (arising due to non-metricity) modifies the kinematics in the averaged universe. This work emphasises this by demonstrating its effects on geometric flows, which in turn affects the distance measures in cosmology.
	
	We perform a 1+3 decomposition of the macroscopic geometry. This enables us to derive the kinematics of geodesic flows. We show that both the Raychaudhuri and Sachs optical equations have contributions from the non-metricity. In other words, back-reaction in macroscopic gravity modifies not only the dynamics but also the kinematics. However, it turns out that the non-metricity has no effect on the cross sectional area of geodesic bundles. We use the Sachs equation to derive an expression for the angular distances in an arbitrary space-time. As a quick application of this geometric setup, we present the case of a spatially flat Friedman-Lema\^itre-Robertson-Walker (FLRW) universe and discuss how the non-metricity affects the distance measurements. We also show that the spatial hypersurfaces have a non-Riemannian geometry and derive the modified Gauss and the Codazzi equations.
	
	The organisation of the paper is as follows: in section \ref{sec-mg}, we present the basic formalism of macroscopic gravity and reinterpret it as a metric-affine theory. In section \ref{sec-kine}, we derive the structure equations for geometric flows. Then, in section \ref{sec-hyper}, we discuss the geometry of spatial hypersurfaces and derive the Gauss-Codazzi equations. In section \ref{sec-cosm}, we derive an expression for the angular distances. We show that the back-reaction affects the distances only through the dynamics (Hubble expansion). Finally, in section \ref{sec-disc}, we discuss the results in this paper and give a few concluding remarks.
	
	The notation used is as follows: Greek indices run from $0$ to $3$ while Latin indices from $1$ to $3$. The partial derivative is denoted by $\boldsymbol{\partial}$ while $\boldsymbol{\nabla}$ and $\boldsymbol{\bar{\nabla}}$ denote the covariant derivatives with respect to the non-Riemannian and the Levi-Civita connections, respectively. Averaged quantities are denoted using angular brackets, $\la \dots \ra$. The indices within $(\dots)$ and $[\dots]$ are are symmetrised/anti-symmetrised; and underlined indices are not included in (anti-)symmetrisation. The sign convention followed is the `Landau-Lifshitz Space-like Convention (LLSC)' \cite{mtw}.

	\section{Averaging the (Pseudo-)Riemannian Geometry} \label{sec-mg}
	\subsection{Basics  of Macroscopic Gravity Formalism} 
	In this section we briefly present the basic formalism of macroscopic gravity. We use the same terminology as used in \cite{zala1,zala2,zala3,zala6}. We will reinterpret certain quantities within the macroscopic gravity formalism in terms of more familiar geometric objects in a metric-affine gravity. Macroscopic gravity employs a covariant averaging procedure to average out the (pseudo-)Riemannian geometry and the field equations of general relativity \cite{zala1,zala2,zala3,zala4,zala5,zala6,zala7}. Given a geometric object, $ p^\alpha_\beta(x) $, defined on an $ n $-dimensional differentiable metric manifold $ (\mathcal{M}, \boldsymbol{g}) $, the space-time averaged value of this object is defined as,
	\begin{equation} \label{mgavg}
		\la p^\alpha_\beta(x) \ra = \dfrac{\int_\Sigma \mathcal{A}^\alpha_{\mu^\prime}(x,x^\prime) p^{\mu^\prime}_{\nu^\prime}(x^\prime) \mathcal{A}^{\nu^\prime}_\beta(x^\prime, x) \sqrt{-\cg^\prime}\ \df^n x^\prime}{\int_\Sigma \sqrt{-\cg^\prime}\ \df^n x^\prime}
	\end{equation}
	where, the average has been taken over a compact region $ \Sigma \subset \mathcal{M} $ around a supporting point $ x \in \Sigma $, $ \int_\Sigma \sqrt{-\cg^\prime}\ \df^n x^\prime $ is the volume ($ V_\Sigma $) of the region $ \Sigma $ and $ \cg^\prime = \det\left[\cg_{\alpha\beta}(x^\prime)\right] $. The integration is done over all the points $ x^\prime \in \Sigma $. The integrand $ \mathcal{A}^\alpha_{\mu^\prime}(x,x^\prime) p^{\mu^\prime}_{\nu^\prime}(x^\prime) \mathcal{A}^{\nu^\prime}_\beta(x^\prime, x) $ is called the bilocal extension of the object $ p^\alpha_\beta(x) $; and the objects $ \mathcal{A}^\alpha_{\mu^\prime}(x,x^\prime) $ and $ \mathcal{A}^{\nu^\prime}_\beta(x^\prime, x) $ are called the bilocal averaging operators. 
	
	The average is taken over finite volume regions in the manifold (i.e., $V_\Sigma$ is finite). Hence, well defined microscopic objects average out without the integral diverging in equation \eqref{mgavg}. In the limit of vanishing averaging volume, the average of the object is equal to the object itself \cite{zala1,zala2,zala3}. Further, there is freedom to choose these averaging operators according to the problem at hand. However, all such operators must follow the following properties: coincidence limit, idempotency, existence of an inverse, and factorisation (see \cite{zala1,zala2} for details). Moreover, the averaging procedure is constructed in such a way that the average of a quantity is uniquely defined and does not depend on the choice of the averaging operators \cite{zala1,zala2,zala3}.
	
	The average of the microscopic Riemann curvature tensor is written as $ \la{R^\alpha}_{\beta\rho\sigma}\ra $ and behaves like a curvature tensor \cite{zala1,zala2,zala3,zala6}. Further, bilocal objects defined as, $ {\mathcal{F}^\alpha}_{\beta\rho}(x,x^\prime) \equiv  \mathcal{A}^\alpha_{\epsilon^\prime}\left(\partial_\rho\mathcal{A}^{\epsilon^\prime}_\beta \right. + \left. \nabla_{\sigma^\prime}\mathcal{A}^{\epsilon^\prime}_\beta\mathcal{A}^{\sigma^\prime}_\rho\right) $, behave as connection coefficients at $ x $ and hence, can be considered to be the bilocal extension of the microscopic connection coefficients, $ {\Gamma^\alpha}_{\beta\rho} $. They have the following coincidence limit,
	\begin{equation}\label{Fcoinlim}
		\lim_{x^\prime\to x} {\mathcal{F}^\alpha}_{\beta\rho}  =  {\Gamma^\alpha}_{\beta\rho} 
	\end{equation} 
	The average of the connection coefficients, $ \la{\Gamma^\alpha}_{\beta\rho}\ra = \frac{\int {\mathcal{F}^\alpha}_{\beta\rho} \sqrt{-\cg^\prime}\ \df^n x^\prime}{\int \sqrt{-\cg^\prime}\ \df^n x^\prime}$, are the macroscopic affine connection coefficients \cite{zala1,zala2}. Then, one cam construct a curvature tensor, $ {M^\alpha}_{\beta\rho\sigma} $, with these macroscopic connection coefficients. This will not be the same as the average of the microscopic curvature tensor, but rather is related to it through,	\begin{equation}\label{macrocurv}
		{M^\alpha}_{\beta\rho\sigma} = \la {R^\alpha}_{\beta\rho\sigma} \ra + 2\la{\mathcal{F}^\delta}_{\beta[\rho}{\mathcal{F}^\alpha}_{\underline{\delta}\sigma]}\ra - 2\la{\mathcal{F}^\delta}_{\beta[\rho}\ra\la{\mathcal{F}^\alpha}_{\underline{\delta}\sigma]}\ra
	\end{equation}
	The additional terms on the right hand side are arising from the fact that the curvature is a non-linear function of the connection coefficients.
	
	One can calculate another connection, $ {\Pi^\alpha}_{\beta\rho} $, associated with the curvature tensor $ \la{R^\alpha}_{\beta\rho\sigma}\ra $. Therefore, in the macroscopic space-time, there are two connections associated with two curvature tensors which are related to each other by equation \eqref{macrocurv}. An affine deformation tensor, $ {A^\alpha}_{\beta\rho} = \la{\Gamma^\alpha}_{\beta\rho}\ra - {\Pi^\alpha}_{\beta\rho}$, can be defined that measures the difference in the two connections. Such a deformation tensor exists due to the metric incompatibility of the connection, $ {\Pi^\alpha}_{\beta\rho} $. That is, for macroscopic metric, $g_{\alpha\beta}$, we have,
	\begin{equation}\label{nonmet1}
		g_{\alpha\beta\lvert\rho} \ne 0
	\end{equation}
	where $ \lvert $ denotes the covariant derivative with respect to the connection $ {\Pi^\alpha}_{\beta\rho} $.
	
	Further, looking at the construction of both the curvature tensors in terms of their respective connections, equation \eqref{macrocurv} takes the form,
	\begin{equation}\label{curvdef}
		{A^\alpha}_{\beta[\rho\|\sigma]} - {A^\epsilon}_{\beta[\rho}{A^\alpha}_{\underline{\epsilon}\sigma]} = -\hf{P^\alpha}_{\beta\rho\sigma}
	\end{equation}
	where $\|$ denotes the covariant derivative with respect to the metric-compatible (Levi-Civita) connection, $\boldsymbol{\la \Gamma \ra}$, of the macroscopic manifold. The tensor, $\boldsymbol{P}$, is dubbed as the gravitational polarisation tensor \cite{zala1}, and is related to the the bilocal objects $\boldsymbol{\mathcal{F}}$ as,
	\begin{equation}\label{curvdef1}
		{P^\alpha}_{\beta\rho\sigma} = 2\la{\mathcal{F}^\delta}_{\beta[\rho}{\mathcal{F}^\alpha}_{\underline{\delta}\sigma]}\ra - 2\la{\mathcal{F}^\delta}_{\beta[\rho}\ra\la{\mathcal{F}^\alpha}_{\underline{\delta}\sigma]}\ra
	\end{equation}	
	The tensor, $ \boldsymbol{P} $, although not a curvature tensor, follows the algebraic and differential properties of one. From equations \eqref{macrocurv} and \eqref{curvdef1}, we can see the that it measures the difference between the two curvature tensors -- a curvature deformation tensor,
	\begin{equation}\label{curvdef2}
		{P^\alpha}_{\beta\rho\sigma} = {M^\alpha}_{\beta\rho\sigma} - \la{R^\alpha}_{\beta\rho\sigma}\ra
	\end{equation}
	
	Both the curvature tensors, $\boldsymbol{M}$ and $\boldsymbol{\la R\ra}$, follow the same algebraic identities and also the differential Bianchi identities with respect to their corresponding connections. The former ensures that the connection, $\Pi$, even though metric incompatible, is symmetric. Moreover, averaging of the differential Bianchi identities for the microscopic curvature tensor gives the following relations,
	\begin{equation}\label{bianchiavg}
		\la{R^\alpha}_{\beta[\rho\sigma}\ra_{,\epsilon]} + \la{R^\gamma}_{\beta[\rho\sigma}{\mathcal{F}^\alpha}_{\underline{\gamma}\epsilon]}\ra - \la{R^\alpha}_{\gamma[\rho\sigma}{\mathcal{F}^\gamma}_{\underline{\beta}\epsilon]}\ra = 0
	\end{equation}	
	In order to evaluate the above equations, a splitting rule is required for the products in the second and third terms. This is constructed using an object called the connection correlation, which is defined as \cite{zala1,zala2},
	\begin{equation}\label{corr2form}
		{Z^\alpha}_{\beta[\gamma}{^\mu}_{\underline{\nu}\sigma]} = \la {\mathcal{F}^\alpha}_{\beta[\gamma}{\mathcal{F}^\mu}_{\underline{\nu}\sigma]} \ra - \la {\mathcal{F}^\alpha}_{\beta[\gamma}\ra \la{\mathcal{F}^\mu}_{\underline{\nu}\sigma]} \ra
	\end{equation} 
	Using equations \eqref{curvdef1} and \eqref{corr2form}, we see that,
	\begin{equation}\label{curvdefcorr2form}
		{P^\alpha}_{\beta\rho\sigma} = 2{Z^\epsilon}_{\beta[\rho}{^\alpha}_{\underline{\epsilon}\sigma]}
	\end{equation}
	
	In summary, when the averaging procedure of macroscopic gravity \eqref{mgavg} is applied to the 4-dimensional (pseudo-)Riemannian manifold of GR, one finds that at macroscopic scales, the manifold is characterised by a metric, two symmetric affine connections{\footnote Gravity theories with multiple connections both in the Riemannian and non-Riemannian framework are already present in the literature \cite{khos,iosbi}.}, two curvature tensors associated with these connections and a correlation tensor that needs to be constructed in order to solve differential Bianchi identities for these curvature tensors. 
	
	\subsubsection{The Field Equations of Macroscopic Gravity}
	The average of Einstein field equations with a cosmological constant,
	\begin{equation}\label{avgefe}
		\la {R^\alpha}_\beta\ra - \hf\delta^\alpha_\beta\la R\ra = \kappa\la {T^\alpha}_\beta\ra - \Lambda \delta^\alpha_\beta
	\end{equation}
	yields the field equations of macroscopic gravity, which are given by \cite{zala1,zala2},
	\begin{equation}\label{macroefe}
		{\bar{R}^\alpha}_\beta - \hf\delta^\alpha_\beta \bar{R} = \kappa\la {T^\alpha}_\beta\ra - \Lambda \delta^\alpha_\beta + \left({Z^\alpha}_{\beta} - \hf\delta^\alpha_\beta P\right) 
	\end{equation}
	where, ${Z^\epsilon}_{\gamma} = {Z^{\epsilon\mu}}_{\mu\gamma}$, $ {Z^\epsilon}_{\mu\nu\gamma} = 2{Z^\epsilon}_{\mu[\alpha}{^\alpha}_{\underline{\nu}\gamma]} $, $P = {P^\mu}_\mu$, $ P_{\mu\nu} = {P^\epsilon}_{\mu\epsilon\nu} $, and $\la \boldsymbol{T}\ra$ is the averaged energy-momentum tensor (usually taken to be that of a perfect fluid in cosmology). Therefore, upon averaging, the field equations get modified and now include additional terms given (partially) in terms of non-metricity of the space-time.

	\subsection{Non-metricity of the Averaged Geometry}
	As discussed above, applying the averaging procedure of macroscopic gravity to GR leads to the interesting result that at macroscopic scales, the manifold is non-Riemannian. That is, the manifold has a Riemannian structure only at small (microscopic) scales. At such scales, GR is a valid theory of gravity. However, when one is concerned with large scales (like in cosmology), an averaging has to be performed on the geometry, and in the process, the Riemannian structure of the manifold is broken. In other words, for cosmological analyses, one must work with a theory of gravity that employs a non-Riemannian geometrical framework. The theory of macroscopic gravity is precisely such a theory. 
	
	According to the macroscopic gravity formalism, the non-Riemannian (macroscopic) manifold is characterised by a symmetric but metric-incompatible connection, given by $\boldsymbol{\Pi}$. The covariant derivative of the metric (of the macroscopic manifold) with respect to this connection is not zero. The non-metricity tensor is then defined as,
	\begin{equation}
		Q_{\rho\alpha\beta} = \hf\nabla_\rho g_{\alpha\beta}
	\end{equation}
	Using this, the general connection, $\boldsymbol{\Pi}$, can be decomposed as,
	\begin{equation}\label{distor1}
		{\Pi^{\rho}}_{\alpha\beta} = {\bar{\Gamma}^\rho}_{\alpha\beta} + {D^{\rho}}_{\alpha\beta} 
	\end{equation}
	where, $ \boldsymbol{\bar{\Gamma}} $ is the Riemannian part (Levi-Civita connection)\footnote{The Levi-Civita connection can be written in terms of metric as,
		\begin{equation}\label{levicivita}
			{\bar{\Gamma}^\rho}_{\alpha\beta} = \hf g^{\rho\sigma}\left(\partial_\alpha g_{\beta\sigma} + \partial_\beta g_{\sigma\alpha} - \partial_\sigma g_{\alpha\beta}\right)
	\end{equation}}, and $ \boldsymbol{D} $ is the so-called disformation tensor that encapsulates the deviation from the Riemannian nature of geometry. The disformation is given in terms of the non-metricity as,
	\begin{equation}\label{qcomb}
		{D^{\rho}}_{\alpha\beta} := {Q^{\rho}}_{\alpha\beta} - {Q_\alpha}^\rho\! _\beta  - {Q_\beta}^\rho\! _\alpha
	\end{equation}
	
	Further, the Ricci identity is given by,
	\begin{equation}\label{ricciidenma}
		\left({\nabla}_\alpha{\nabla}_\beta - {\nabla}_\beta{\nabla}_\alpha \right)V^\rho = {{R}^\rho}_{\sigma\alpha\beta}V^\sigma
	\end{equation}
	where, $ {{R}^\rho}_{\sigma\alpha\beta} $, is the Riemann curvature tensor defined in terms of the general connection of the non-Riemannian geometry,
	\begin{equation}\label{riemcurvma}
		{{R}^\rho}_{\sigma\alpha\beta} := \partial_\alpha {{\Pi}^\rho}_{\sigma\beta} - \partial_\beta {{\Pi}^\rho}_{\sigma\alpha} + {{\Pi}^\rho}_{\mu\alpha}{{\Pi}^\mu}_{\sigma\beta} - {{\Pi}^\rho}_{\mu\beta}{{\Pi}^\mu}_{\sigma\alpha}
	\end{equation}
	Using equation \eqref{distor1}, it can be shown that,
	\begin{equation}\label{curvdecomp}
		{{R}^\rho}_{\sigma\alpha\beta} = {\bar{R}^\rho}_{\sigma\alpha\beta} + \bar{\nabla}_\alpha {{D}^\rho}_{\sigma\beta} - \bar{\nabla}_\beta {{D}^\rho}_{\sigma\alpha} + {{D}^\rho}_{\mu\alpha}{{D}^\mu}_{\sigma\beta} - {{D}^\rho}_{\mu\beta}{{D}^\mu}_{\sigma\alpha} 
	\end{equation}
	where, $ {\bar{R}^\rho}_{\sigma\alpha\beta} $ is the Riemannian part of the curvature tensor which is the same as $\boldsymbol{M}$ of the previous section, and $\bar{\nabla}$ is the covariant derivative both defined with the Levi-Civita connection. One can define a new quantity that captures the deviation of the curvature tensor from its Riemannian counterpart,
	\begin{equation}
		{{N}^\rho}_{\sigma\alpha\beta} := \bar{\nabla}_\alpha {{D}^\rho}_{\sigma\beta} - \bar{\nabla}_\beta {{D}^\rho}_{\sigma\alpha} + {{D}^\rho}_{\mu\alpha}{{D}^\mu}_{\sigma\beta} - {{D}^\rho}_{\mu\beta}{{D}^\mu}_{\sigma\alpha}
	\end{equation}
	
	It is not difficult to see that the so-called affine deformation tensor, $\boldsymbol{A}$, and the polarisation tensor, $\boldsymbol{P}$, of macroscopic gravity are, respectively, the disformation tensor, $\boldsymbol{D}$, and the object, $\boldsymbol{N}$, of the metric-affine gravity literature. Henceforth, we will use these symbols to denote the affine deformation tensor and polarisation tensor since we are treating the theory of macroscopic gravity as a metric-affine theory.

	\section{Structure Equations for Geometric Flows} \label{sec-kine}
	Since the geometry at the macroscopic scales is no longer Riemannian, determining the geodesic curves (curves with extremum length) is not trivial. In Riemannian geometry, the autoparallel curves with respect to the Levi-Civita connection and also the geodesic curves. Then, the geodesic equation is given by,
	\begin{equation}\label{geodesiceq}
		u^\beta \bar{\nabla}_\beta u^\alpha = 0
	\end{equation}
	
	However, in general, autoparallel curves (of a non-Riemannian geometry) are not necessarily geodesics. In fact, since in non-Riemannian geometry, there exist additional geometric fields (like non-metricity) the paths traced by freely moving physical particles (the geodesics) will depend on whether these particles interact with these `new' geometric fields \cite{obu}. In cosmology, one takes the content of the universe to be a perfect fluid or dust. Since such matter does not interact with non-metricity, despite the averaged geometry being non-Riemannian, the path traced is taken to be Riemannian geodesics given by equation \eqref{geodesiceq}.
	
	We will now derive equations governing the evolution of the congruences of geodesics on the averaged (non-Riemannian) geometry. These are sometimes called kinematic equations. Given a congruence, one can characterise its flow in terms of the evolution of the irreducible components of its velocity gradient. Physically, these equations characterise the flow of the fluid content of the Universe (for the timelike curves) and bundle of photons (for the null curves). We will derive (in detail) the equations for the trace part of the velocity gradient (also called expansion) for both the timelike and null curves. This will give us the macroscopic gravity analogue of the Raychaudhuri and Sachs equations \cite{ray1,sachs1}.
	
	\subsection{The Raychaudhuri Equation}
	To derive the Raychaudhuri equation, we begin with a congruence of timelike geodesics. Let $u^\alpha$ be the vector field tangent to the congruence (the velocity vector). In cosmology, this is simply the comoving velocity of the (fluid) content of the Universe, $u^\alpha = (1, 0,0,0)$. The norm of such a velocity is normalised as, $u^\alpha u_\alpha = -1$ \cite{ioscos}. We are interested in the transverse evolution of the congruence and hence we define a transverse metric in the following manner \cite{ioscos},
	\begin{equation}\label{indmet}
		h_{\alpha\beta} = g_{\alpha\beta} + u_\alpha u_\beta
	\end{equation}
	The 3-metric, $h_{\alpha\beta}$, is completely transverse to the congruence, i.e., we have, $u^\alpha h_{\alpha\beta} = 0 = h_{\alpha\beta}u^\beta$. It is easy to see that, ${h^\alpha}_\alpha = 3$. This splits the space-time in temporal and spatial parts, referred to as $1+3$ decomposition. All the quantities transverse to the congruence reside on a spatial hypersurface orthogonal to it. Here, we will investigate the transverse properties of the congruence while in the next section we present the properties of spatial hypersurfaces.
	
	Using the 3-metric, one can separate out the transverse and longitudinal parts of the velocity gradient as follows,
	\begin{align}
		^{{\rm (T)}}\nabla_\alpha u^\beta &= \tensor{h}{^\rho_\alpha} \tensor{h}{^\beta _\epsilon} \nabla_\rho u^\epsilon \label{transvelgrad}\\
		^{{\rm (L)}}\nabla_\alpha u^\beta &= \nabla_\alpha u^\beta - ^{{\rm (T)}}\nabla_\alpha u^\beta \label{longvelgrad}
	\end{align}
	where, $\nabla$ represents the covariant derivative with respect to the non-Riemannian connection, $\boldsymbol{\Pi}$. It is easy to check that, $u^\alpha\ ^{{\rm (T)}}\nabla_\alpha u^\beta = 0 = ^{{\rm (T)}}\nabla_\alpha u^\beta u_\beta$.
	
	The velocity gradient can be decomposed into its irreducible components as follows,
	\begin{align} \label{kinedef}
		\theta &:= \delta^\alpha_\beta\ ^{{\rm (T)}}\nabla_\alpha u^\beta = {h^\alpha}_\beta \nabla_\alpha u^\beta \\
		{\omega_\alpha}^\beta &:= ^{{\rm (T)}}\nabla_{[\alpha} u^{\beta]} = {h^\rho}_{[\alpha}{h^{\beta]}}_\epsilon \nabla_\rho u^\epsilon \\
		{\sigma_\alpha}^\beta &:= ^{{\rm (T)}}\nabla_{(\alpha} u^{\beta)} - \frac{1}{3} \theta {h_\alpha}^\beta = {h^\rho}_{(\alpha}{h^{\beta)}}_\epsilon \nabla_\rho u^\epsilon - \frac{1}{3} \theta {h_\alpha}^\beta
	\end{align}
	where, $\theta$ is the trace part known as expansion, $\boldsymbol{\omega}$ is the anti-symmetric part known as rotation or vorticity, and $\boldsymbol{\sigma}$ is the symmetric traceless part known as shear.
	
	Using the expression for the expansion scalar, we can write,
	\begin{equation}
		\theta = \nabla_\alpha u^\alpha + D_{\alpha\beta\rho} u^\alpha u^\beta u^\rho 
	\end{equation} 
	where, we have used, $u^\alpha \nabla_\alpha u^\beta = \tensor{D}{^\beta_\alpha_\rho}u^\alpha u^\rho$ which follows from equation \eqref{geodesiceq}. One can decompose the first term above into a Riemannian and a non-Riemannian part. The Riemannian part would be the usual expansion scalar, $\bar{\theta} = \bar{\nabla}_\alpha u^\alpha$. Then, we have,
	\begin{equation}
		\theta = \bar{\theta} + \tensor{D}{^\alpha_\alpha_\beta}u^\beta + D_{\alpha\beta\rho} u^\alpha u^\beta u^\rho
	\end{equation}
	Taking a derivative on both sides gives us the Raychaudhuri equation in the averaged geometry,
	\begin{equation}
		\frac{\df \theta}{\df \lambda} = -\bar{R}_{\alpha\beta}u^\alpha u^\beta - \frac{1}{3}\bar{\theta}^2 + \tensor{\bar{\omega}}{^\alpha_\beta} \tensor{\bar{\omega}}{_\alpha^\beta} - \tensor{\bar{\sigma}}{^\alpha_\beta} \tensor{\bar{\sigma}}{_\alpha^\beta} + \frac{\df}{\df \lambda}\left(\tensor{D}{^\alpha_\alpha_\beta}u^\beta + D_{\alpha\beta\rho} u^\alpha u^\beta u^\rho\right)
	\end{equation}
	where, the terms with a bar over them are Riemannian quantities. The last term on the right hand side is the contribution from averaging (the back-reaction) given solely on terms of the non-metricity of the averaged space-time. 
	
	\subsection{The Sachs Optical Equation}
	The Sachs optical equation is the null counterpart of the Raychaudhuri equation. To derive it, we begin with considering a congruence of null geodesics with tangent vector, $k^\alpha$. The norm of the tangent is, $k^\alpha k_\alpha = 0$. In this case, the transverse metric is given by,
	\begin{equation}
		H_{\alpha\beta} = g_{\alpha\beta} + k_\alpha n_\beta + n_\alpha k_\beta
	\end{equation}
	where, $n^\alpha$ is an auxiliary null vector satisfying the conditions, $n^\alpha n_\alpha = 0$ and $n^\alpha k_\alpha = -1$. Once again, one can check, $k^\alpha H_{\alpha\beta} = 0 = H_{\alpha\beta}k^\beta$ and ${h^\alpha}_\alpha = 2$. 
	
	The steps involved in deriving the Sachs equation are analogous to previous section. We define the transverse part of the velocity gradient as,
	\begin{align}
		^{{\rm (T)}}\nabla_\alpha k^\beta = \tensor{H}{^\rho_\alpha} \tensor{H}{^\beta _\epsilon} \nabla_\rho k^\epsilon \label{transvelgradnull}
	\end{align}
	This can be can be decomposed into irreducible components as follows,
	\begin{align} \label{kinedefnull}
		\Theta &:= \delta^\alpha_\beta\ ^{{\rm (T)}}\nabla_\alpha k^\beta = {H^\alpha}_\beta \nabla_\alpha k^\beta , \\
		{\Omega_\alpha}^\beta &:= ^{{\rm (T)}}\nabla_{[\alpha} k^{\beta]} = {H^\rho}_{[\alpha}{H^{\beta]}}_\epsilon \nabla_\rho k^\epsilon, \\
		{\Sigma_\alpha}^\beta &:= ^{{\rm (T)}}\nabla_{(\alpha} k^{\beta)} - \frac{1}{3} \Theta {H_\alpha}^\beta = {H^\rho}_{(\alpha}{H^{\beta)}}_\epsilon \nabla_\rho k^\epsilon - \frac{1}{3} \Theta {H_\alpha}^\beta,
	\end{align}
	Using the definition of the expansion scalar, we get,
	\begin{equation}\label{expnull}
		\Theta = \nabla_\alpha k^\alpha + D_{\alpha\beta\rho}n^\alpha k^\beta k^\rho + D_{\alpha\beta\rho}k^\alpha k^\beta n^\rho
	\end{equation}
	where, we have used, $k_\beta \nabla_\alpha k^\beta = D_{\alpha\beta\rho}k^\alpha k^\beta$. The first term can again be decomposed into Riemannian and non-Riemannian parts. The Riemannian part is the usual null expansion scalar, $\bar{\Theta} = \bar{\nabla}_\alpha k^\alpha$. This gives us,
	\begin{equation}\label{expnull1}
		\Theta = \bar{\Theta} + {D^\alpha}_{\alpha\beta}k^\beta + D_{\alpha\beta\rho}n^\alpha k^\beta k^\rho + D_{\alpha\beta\rho}k^\alpha k^\beta n^\rho
	\end{equation}
	Taking a derivative on both sides gives the Sachs optical equation in the averaged geometry,
	\begin{equation}\label{sachseq}
		\frac{\df \Theta}{\df \lambda} = -\bar{R}_{\alpha\beta}k^\alpha k^\beta - \frac{1}{2}\bar{\Theta}^2 + \tensor{\bar{\Omega}}{^\alpha_\beta} \tensor{\bar{\Omega}}{_\alpha^\beta} - \tensor{\bar{\Sigma}}{^\alpha_\beta} \tensor{\bar{\Sigma}}{_\alpha^\beta} + \frac{\df}{\df \lambda}\left({D^\alpha}_{\alpha\beta}k^\beta + D_{\alpha\beta\rho}n^\alpha k^\beta k^\rho + D_{\alpha\beta\rho}k^\alpha k^\beta n^\rho\right)
	\end{equation}	
	
	Therefore, the expansion of both the timelike and null geodesics is affected by the non-metricity that arises due to averaging. The non-metricity in macroscopic gravity partially constitutes cosmological back-reaction. Hence, we have shown here that the back-reaction not only modifies the dynamics but also the kinematics in the averaged Universe. This is expected since in addition to the field equations, the process of averaging in macroscopic gravity also modifies the underlying geometric structure.

	\subsubsection{Expansion Scalar and Cross Section of Null Bundles}\label{sec-exparearel}
	It is useful (especially in cosmology) to show how the expansion scalar is related to the cross sectional area of the congruence. A formal way of doing this can be found in \cite{poisson}. We will follow an analogous treatment here. Consider a cross section around a point $P$ on a curve in the congruence. This can be constructed by taking a set of points, $P^\prime$, on the neighbouring curves that correspond to the same parameter value as that of $P$. Let this set of points be called, $S_P$. To see the change in cross section, we need to compare $S_P$ to a similar set of neighbouring points at some other point, $Q$, along the same curve. Let us call this $S_Q$. Since the points on $S_P$ (and $S_Q$) are also points on neighbouring curves, they can be used to label the curves, thus giving us a consistent coordinate system, $y^A\ (A = 2,3)$, on the cross section. Then, one can construct a 2-metric on the cross section as,
	\begin{equation}
		\df s^2 = H_{AB}\ \df y^A \df y^B
	\end{equation} 
	
	Using this, it can be shown that (see \cite{poisson}) the fractional rate of change of the cross sectional area is,
	\begin{equation}
		\frac{1}{A}\frac{\df A}{\df \lambda} = \hf H^{AB} \frac{\df H_{AB}}{\df \lambda}
	\end{equation}
	Calculating the right hand side of the above equation, we get,
	\begin{equation}
		\hf H^{AB} \frac{\df H_{AB}}{\df \lambda} = \nabla_\alpha k^\alpha - {D^\alpha}_{\alpha\beta}k^\beta
	\end{equation}
	Using equation \eqref{expnull1}, this becomes,
	\begin{equation}\label{crosssec}
		\hf H^{AB} \frac{\df H_{AB}}{\df \lambda} = \bar{\Theta}
	\end{equation}
	Therefore, the fractional rate of change of the cross sectional area of the null geodesic congruence in the averaged geometry is equal to the Riemannian part of the expansion scalar,
	\begin{equation}\label{exparearel}
		\bar{\Theta} = \frac{1}{A}\frac{\df A}{\df \lambda}
	\end{equation}
	An important point to note here is that this is turning out to be the case since we are dealing with geodesic curves ($k^\alpha\bar{\nabla}_\alpha k^\beta = 0$). For non-geodesic curves one should expect contributions from non-metricity \footnote{In an even more general metric-affine geometry where the connection is not symmetric, one would have contributions from torsion even in case of geodesic curves \cite{aga1}.}. The fractional rate of change of cross sectional area is closely related to the angular diameter distances in cosmology. The consequence of equation \eqref{crosssec} is that the non-metricity will contribute to the distances only through the dynamics.

	\section{The Geometry of Hypersurfaces}\label{sec-hyper}
	In this section, we will continue with the $1+3$ decomposition of the space-time presented in the last section. However, here we will present the equations governing the geometric properties of the spatial hypersurfaces. This approach is commonly employed in GR to split the curvature and the field equations in temporal and spatial parts. This leads to the well known Gauss-Codazzi equations and the definitions of the shift and lapse functions. Such a splitting is particularly useful in probing various physical situations in numerical relativity.
	
	Let us consider a spacelike hypersurface normal to a timelike geodesic. The 3-metric, $h_{\alpha\beta}$, in equation \eqref{indmet} serves as the metric on this hypersurface. This is sometimes referred to as the induced metric. In the signature adapted in this paper, this is a positive definite non-Riemannian metric. To analyse the geometric properties of the hypersurface, one needs to take the spatial projection of the space-time objects. This is done using the induced metric in a similar manner as equation \eqref{transvelgrad}.
	
	\subsection{The Non-Metricity of the Hypersurface}
	The spatial projection of the covariant derivative is defined as,
	\begin{equation}
		^{(3)}\nabla_\mu {T^{\alpha_1 \alpha_2\dots}}_{\beta_1\beta_2\dots} := {h^{\alpha_1}}_{\rho_1} {h^{\alpha_2}}_{\rho_2} \dots {h^{\epsilon_1}}_{\beta_1} {h^{\epsilon_2}}_{\beta_2}  \dots {h^{\nu}}_{\mu} \nabla_\nu {T^{\rho_1 \rho_2\dots}}_{\epsilon_1\epsilon_2\dots}
	\end{equation}
	This serves as the covariant derivative on the hypersurface. Using this definition, we can write,
	\begin{equation}
		^{(3)}\nabla_\rho h_{\alpha\beta} = {h^\epsilon}_{\rho}{h^\mu}_\alpha{h^\nu}_\beta \nabla_\epsilon h_{\mu\nu} = 2{h^\epsilon}_{\rho}{h^\mu}_\alpha{h^\nu}_\beta Q_{\epsilon\mu\nu}
	\end{equation}
	Therefore, we see that the spatial covariant derivative of the 3-metric is not zero. In other words, the spatial connection is not metric compatible\footnote{This could mean that a covariant averaging, albeit over 3-space, necessarily breaks the Riemannian nature of the space. One needs to be careful about this when comparing macroscopic gravity to spatial averaging schemes that deal only with scalars.}. This means that the hypersurface retains the non-Riemannian nature. We define the the non-metricity of the hypersurface as,
	\begin{equation}
		^{(3)}Q_{\rho\alpha\beta} := \hf\ ^{(3)}\nabla_\rho h_{\alpha\beta} = {h^\epsilon}_{\rho}{h^\mu}_\alpha{h^\nu}_\beta Q_{\epsilon\mu\nu}
	\end{equation}
	Similarly, one can evaluate that, $\hf\ ^{(3)}\nabla_\rho h^{\alpha\beta} = -^{(3)}{Q_\rho}^{\alpha\beta}$. Note, however, that, $^{(3)}\nabla_\rho {h^\alpha}_\beta = 0$.
	
	The extrinsic curvature of the hypersurface is defined as,
	\begin{equation}\label{extcurv1}
		K_{\alpha\beta} :=\ ^{(3)} \nabla_\alpha u_\beta
	\end{equation}
	where, $\boldsymbol{u}$ is a timelike vector normal to the hypersurface i.e., $u^\alpha K_{\alpha\beta} = 0 = K_{\alpha\beta}u^\beta$. Due to the non-metricity of the hypersurface, we have,
	\begin{equation}\label{extcur2}
		^{(3)} \nabla_\alpha u^\beta = {K_\alpha}^\beta - 2 {h^\rho}_\alpha {h^\beta}_\epsilon {Q_\rho}^{\epsilon\sigma}u_\sigma
	\end{equation}
	where,	${K_\alpha}^\beta = g^{\rho \beta}K_{\alpha\rho} $. Since the connection is torsion-less (and hence, so is the connection on the hypersurface), we have, $ K_{\alpha\beta} = K_{\beta\alpha} $.
	
	\subsection{The Gauss Equation}
	The Gauss equations arises from evaluating the Ricci identity on the hypersurface. It gives an expression for the Riemannian curvature of the full space-time in terms of its components projected onto the hypersurface. To derive the Gauss equation, we begin with defining the Riemann tensor on the hypersurface in the following manner, 
	\begin{equation}
		\left[^{(3)}\nabla_\alpha\ ^{(3)}\nabla_\beta - ^{(3)}\nabla_\beta\ ^{(3)}\nabla_\alpha\right]n^\rho =\  ^{(3)}{R^\rho}_{\sigma\alpha\beta}n^\sigma
	\end{equation}
	where, $\boldsymbol{n}$ is some vector tangential to the hypersurface, i.e., $n^\alpha u_\alpha = 0$. Evaluating the left side of the above equation, we get,
	\begin{equation}
		^{(3)}{R^\rho}_{\sigma\alpha\beta}n^\sigma = \left[ ^{(3)}\nabla_\alpha u_\sigma\ ^{(3)}\nabla_\beta u^\rho -\ ^{(3)}\nabla_\beta u_\sigma\ ^{(3)}\nabla_\alpha u^\rho \right] n^\sigma + {h^\rho}_\epsilon {h^\mu}_\alpha {h^\nu}_\beta \left(\nabla_\mu \nabla_\nu - \nabla_\nu\nabla_\mu\right)n^\epsilon 
	\end{equation}
	
	Using equations \eqref{ricciidenma}, \eqref{extcurv1}, \eqref{extcur2}, and recognising, $n^\phi = {h^\phi}_\sigma n^\sigma$, we finally get the Gauss equation,
	\begin{equation}
		^{(3)}{R^\rho}_{\sigma\alpha\beta} = {h^\rho}_\epsilon {h^\phi}_\sigma {h^\mu}_\alpha {h^\nu}_\beta{R^\epsilon}_{\phi\mu\nu} + K_{\alpha\sigma}{K^\rho}_\beta - K_{\beta\sigma} {K^\rho}_\alpha - 2 {h^\rho}_\epsilon \tensor{Q}{_\nu ^\epsilon _\mu}u^\mu\left(K_{\alpha\sigma}{h^\nu}_\beta - K_{\beta\sigma}{h^\nu}_\alpha\right)
	\end{equation}
	Contracting the appropriate indices, we can write similar equations for the Ricci tensor and Ricci scalar,
	\begin{equation}
		^{(3)}R_{\sigma\beta} = {h^\phi}_\sigma {h^\nu}_\beta R_{\phi\nu} + K_{\alpha\sigma}{K^\alpha}_\beta - K_{\beta\sigma} {K^\alpha}_\alpha - 2 \tensor{Q}{_\nu ^\epsilon _\mu}u^\mu \left( K_{\epsilon\sigma}{h^\nu}_\beta - K_{\beta\sigma}{h^\nu}_\epsilon \right)
	\end{equation}
	\begin{equation}
		^{(3)} R = R + R_{\alpha\beta}u^\alpha u^\beta + K_{\alpha\beta}K^{\alpha\beta} - \left({K^\alpha}_\alpha\right)^2 - 2 \tensor{Q}{_\alpha ^\beta _\rho}u^\rho \left({K^\alpha}_\beta - {K^\epsilon}_\epsilon {h^\alpha}_\beta\right)
	\end{equation}
	The above identities are useful in performing a $1+3$ splitting of the field equations.
	
	\subsection{The Codazzi Equation}
	The Gauss equation relates the Riemann curvature tensor with all components projected onto the hypersurface to the extrinsic curvature. Similar to this, one can project only three components of the Riemann tensor onto the hypersurface and one component along a timelike vector, $\boldsymbol{u}$. This quantity also be written in terms of the extrinsic curvature. This is done through the Codazzi equation.
	
	To derive the Codazzi equation, we start with evaluating,
	\begin{equation}
		^{(3)}\nabla_\rho K_{\alpha\beta} = - \left(K_{\rho\alpha}{h^\epsilon}_\beta + K_{\rho\beta}{h^\epsilon}_\alpha \right)D_{\mu\nu\epsilon}u^\mu u^\nu + {h^\epsilon}_\rho {h^\mu}_\alpha {h^\nu}_\beta \nabla_\epsilon\nabla_\mu u_\nu
	\end{equation}
	Anti-symmetrising $\rho$ and $\alpha$, the first term in the brackets on the right hand side vanishes. The second term turns into the Ricci identity \eqref{ricciidenma}. Taking care of raising and lowering of the indices introduces a few non-metricity terms, finally giving,
	\begin{equation}
		^{(3)}\nabla_\rho K_{\alpha\beta} -\ ^{(3)}\nabla_\alpha K_{\rho\beta} = {h^\epsilon}_\rho {h^\mu}_\alpha {h^\nu}_\beta \left(R_{\epsilon \sigma \mu\nu} + 4\nabla_{[\epsilon}Q_{\mu]\nu\sigma} \right) u^\sigma + \left(K_{\alpha\beta}{h^\epsilon}_\rho - K_{\rho\beta}{h^\epsilon}_\alpha \right)D_{\mu\nu\epsilon}u^\mu u^\nu 
	\end{equation}
	This is the Codazzi equation. Contracting the indices $\rho$ and $\beta$ in the above equation, we get,
	\begin{multline}
		^{(3)}\nabla_\rho {K^\rho}_\alpha -\ ^{(3)}\nabla_\alpha {K^\beta}_\beta = -{h^\mu}_\alpha R_{\sigma\mu}u^\sigma + {h^\mu}_\alpha {h^\epsilon}_\nu \left( \nabla_{[\epsilon}\tensor{Q}{_\mu_]^\nu_\sigma} + \tensor{Q}{_[_\epsilon^\rho^\nu}\tensor{Q}{_\mu_]_\rho_\sigma} \right) u^\sigma \\+ \left({K^\epsilon}_\alpha - {K^\beta}_\beta{h^\epsilon}_\alpha \right)D_{\mu\nu\epsilon}u^\mu u^\nu -\ 4\ ^{(3)}\tensor{Q}{_( _\rho^\rho^\beta} K_{\alpha)\beta}  
	\end{multline}
	
	Using the $1+3$ covariant formalism presented here, it will be straightforward to decompose the four dimensional averaged geometry into foliations and fibrations as is frequently done. The definition of the lapse function and the shift vector remain the same and the $1+3$ decomposition of the metric retains the same form \cite{cap}.
	
	\section{FLRW Cosmology in Macroscopic Gravity}	\label{sec-cosm}
	The field equations of macroscopic gravity, obtained by averaging the Einstein field equations, contain different traces of the connection correlation and the non-Riemannian part of the curvature tensor. These additional terms in equation \eqref{macroefe} are the result of a change in the geometric structure of the space-time due to averaging. These terms constitute precisely what is referred to as the cosmological back-reaction in the literature. Considering a particular averaged geometry (i.e., taking an ansatz on the metric of the averaged space-time), one can find the back-reaction in terms of the metric coefficients. This has been previously done for the FLRW cosmological model \cite{coley,hoogen2,tharake1}. Here, we will build upon the analyses in \cite{coley,hoogen2,tharake1} and discuss how the non-metricity affects the distance measurements in cosmology. 
	
	\subsection{Dynamics and Expansion History}
	Let us begin by considering the metric at the macroscopic scales is the spatially flat FLRW metric,
	\begin{equation}\label{exactflrw}
		\df s^2 = -\df t^2 + a^2(t)\left(\df r^2 + r^2\df\theta^2 + r^2\sin^2\theta\df \phi^2 \right)
	\end{equation}
	Considering the averaged matter distribution to be a perfect fluid, we have,
	\begin{equation}\label{emtensor}
		\la {T^\epsilon}_\gamma \ra = (\rho + p) u^\epsilon u_\gamma + p \delta^\epsilon_\gamma
	\end{equation}
	where, $\rho \equiv \rho(t)$ and $p \equiv p(t)$ are the energy density and pressure for the averaged matter-energy content, respectively; and $ u^\alpha~=~\left(1,0,0,0\right) $ is the average 4-velocity of the fluid. Then, writing equation \eqref{macroefe} explicitly gives us the modified Friedmann equations \cite{coley,hoogen2},
	\begin{align}
		3\left(\frac{\dot{a}}{a}\right)^2  &= 8\pi \rho +\Lambda - \frac{\mathcal{Q}}{a^2} \label{fried1} \\
		2\frac{a\ddot{a}}{a^2} + \left(\frac{\dot{a}}{a}\right)^2 &= -8\pi p + \Lambda - \frac{\mathcal{Q}}{3a^2} \label{fried2}
	\end{align} 
	where, $\mathcal{Q}$ is a constant coming from the back-reaction term and `dot' represents the derivative with respect to the cosmological time, $t$. One can immediately notice that the back-reaction term evolves as a curvature.
	
	The first equation above can be written in the form,
	\begin{equation}
		H^2 = \frac{8\pi}{3} (\rho_m + \rho_\Lambda + \rho_\mathcal{Q})
	\end{equation}
	where, $H(t) = \frac{\dot{a}}{a}$ is the Hubble parameter, we have defined, $\rho_\Lambda = \frac{\Lambda}{8\pi}$ and $\rho_\mathcal{Q} = -\frac{\mathcal{Q}}{8\pi a^2}$, and we have added the subscript $m$ to differentiate the matter density from other energy densities. Further, defining the associated density parameters as, $\Omega_X = \frac{8\pi \rho_X}{3H^2}$, we can write,
	\begin{equation}
		1 = \Omega_m + \Omega_\Lambda + \Omega_\mathcal{Q} 
	\end{equation}
	Multiplying both sides by a factor of $\frac{H^2}{H_0^2}$, we get,
	\begin{equation}
		H^2 = H_0^2 \left( \Omega_{m_0} a^{-3} + \Omega_{\Lambda_0} + \Omega_{\mathcal{Q}_0} a^{-2} \right) = H_0^2 \left[ \Omega_{m_0} (1+z)^{3} + \Omega_{\Lambda_0} + \Omega_{\mathcal{Q}_0} (1+z)^{2} \right]
	\end{equation}
	where, $\Omega_{X_0} = \frac{8\pi \rho_{X_0}}{3H_0^2}$, $\rho_{\mathcal{Q}_0} = -\frac{\mathcal{Q}}{8\pi}$ is the `non-metricity density' today, and we have used, $a = \frac{1}{1+z}$. The above expression gives the expansion history of the universe taking into account the effect due to non-metricity at the macroscopic scales. One can see that depending on the sign of the constant, $\mathcal{Q}$, a spatially flat FLRW Universe may evolve identical to a spatially curved one.
	
	\subsection{Angular Diameter Distance}
	In observational cosmology, one deals with large amounts photons, and therefore, it is the
	averaged photon paths that should be used to calculate distances. In the context of macroscopic gravity, this means the paths of photon bundles (null geodesics) on the macroscopic geometry should be used to determine the distance measures in cosmology. The angular diameter distance, $D_A$, is related to the cross sectional area, $A$, of null congruence as \cite{peebles},
	\begin{equation}
		A \propto D_A^2
	\end{equation}
	Using this, one can write,
	\begin{equation}
		\frac{1}{A}\frac{\df A}{\df \lambda} = \frac{2}{D_A}\frac{\df D_A}{\df \lambda}
	\end{equation}
	Using equation \eqref{exparearel}, this becomes,
	\begin{equation}
		\bar{\Theta} = \frac{2}{D_A}\frac{\df D_A}{\df \lambda}
	\end{equation}
	Taking a derivative on both sides, one gets,
	\begin{equation}\label{disteq1}
		\frac{2}{D_A}\frac{\df^2 D_A}{\df \lambda^2} = \frac{\df \bar{\Theta}}{\df \lambda} + \hf \bar{\Theta}^2 = -\bar{R}_{\alpha\beta}k^\alpha k^\beta + \tensor{\bar{\Omega}}{^\alpha_\beta} \tensor{\bar{\Omega}}{_\alpha^\beta} - \tensor{\bar{\Sigma}}{^\alpha_\beta} \tensor{\bar{\Sigma}}{_\alpha^\beta}
	\end{equation}
	where, in the second equality we have used the Sachs equation \eqref{sachseq}.
	
\begin{figure}[H]
		\centering
			\includegraphics[width = 0.5\linewidth]{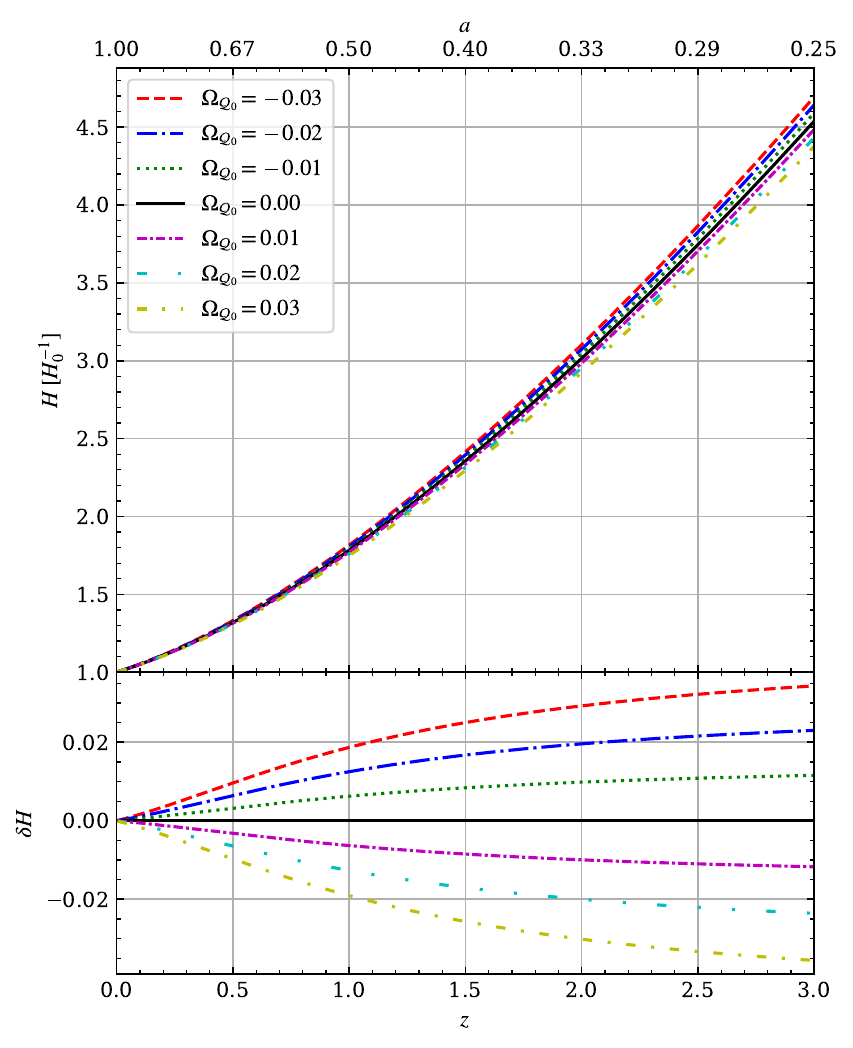} \hspace{-0.35cm}
			\includegraphics[width = 0.5\linewidth]{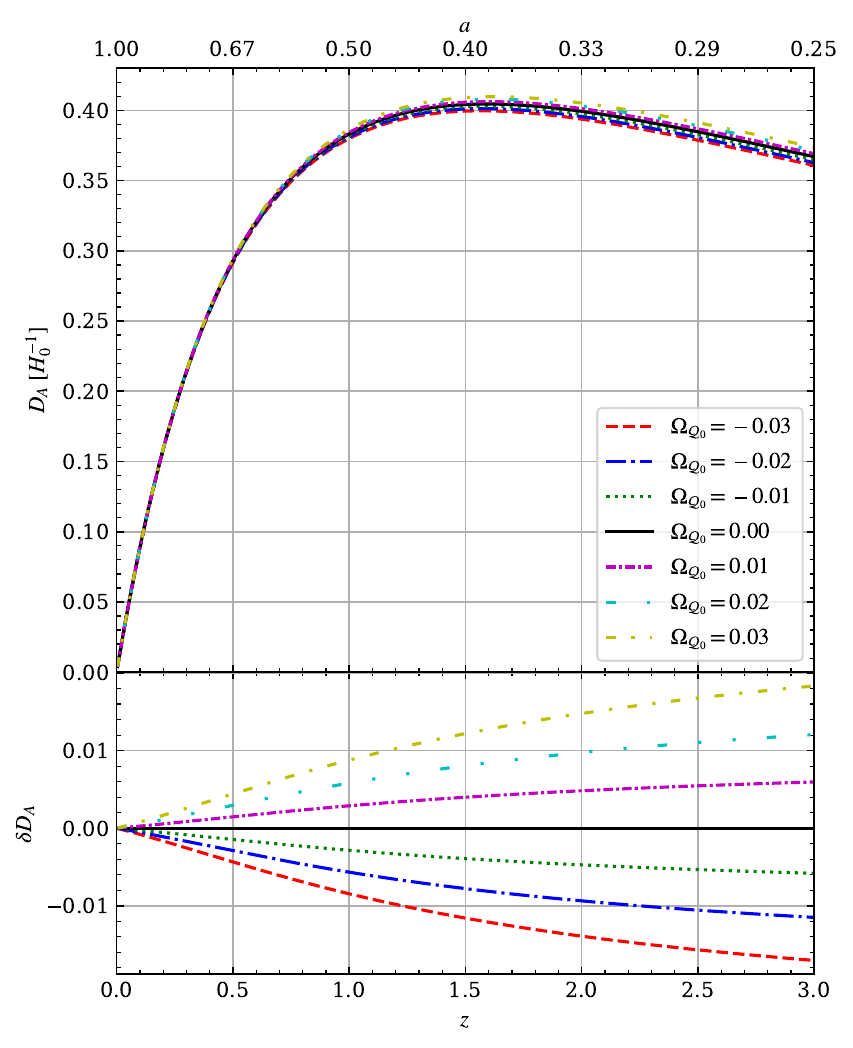} 
		\caption{\small Hubble parameter (\textit{left}) and angular diameter distance (\textit{right}) at late times ($0<z<3$) for small values of the non-metricity parameter ($-0.03\le\Omega_{\mathcal{Q}_0}\le0.03$). The lower panel shows the relative difference between $\Lambda $CDM values and the $\mathcal{Q}$-$\Lambda$CDM values.} \label{fig-plots}
	\end{figure}

	The above equation holds for arbitrary space-times. For a spatially flat FLRW space-time, we have, $\boldsymbol{\bar{\Omega}} = 0 = \boldsymbol{\bar{\Sigma}}$, and the 4-velocity of null geodesics is given by \cite{pleb},
	\begin{equation}
		k^\alpha = \left(-a^{-1}, a^{-2},0,0\right)
	\end{equation}
	Using this, one can calculate,
	\begin{equation}
		-\bar{R}_{\alpha\beta}k^\alpha k^\beta = \frac{2\dot{H}}{a}
	\end{equation}
	Using this in equation \eqref{disteq1}, we get,
	\begin{equation}\label{angdiseqflrw}
		\frac{\df^2 D_A}{\df \lambda^2} -\frac{1}{a^2}\frac{\df H}{\df t} D_A = 0
	\end{equation}
	Using $k^0=\frac{\df t}{\df \lambda}= -\frac{1}{a}$, the above equation becomes,
	\begin{equation}\label{angdiseqt}
		\frac{\df^2 D_A}{\df t^2} - H\frac{\df D_A}{\df t} - \frac{\df H}{\df t} D_A = 0
	\end{equation}
	Writing in terms of the scale factor, $a$, this becomes,
	\begin{equation}\label{andiseqa}
		\frac{\df^2 D_A}{\df a^2} + \frac{1}{H}\frac{\df H}{\df a}\frac{\df D_A}{\df a} - \frac{1}{aH}\frac{\df H}{\df a}D_A = 0
	\end{equation}
	Now, making a variable change, $\chi(a)=\int\frac{\df a}{a^2H}$, we get,
	\begin{equation}\label{angdisv}
		\frac{\df^2 D_A}{\df \chi^2} - 2aH\frac{\df D_A}{\df \chi} - \frac{\df H}{\df \chi}D_A = 0
	\end{equation} 
	which can be reduced to,
	\begin{equation}\label{angdiseqmain}
		\frac{\df^2}{\df \chi^2}\left(\frac{D_A}{a}\right) = 0
	\end{equation}
	Solving the above differential equation leads to an expression for the angular diameter distance in the FLRW geometry,
	\begin{equation}\label{angdisexp}
		D_A(a) = a\chi
	\end{equation}
	Therefore, we see that the non-metricity affects the distance measurements only through its presence in the Hubble parameter. That is, even though the cross sectional area of null congruences only depends on the Riemannian part of the expansion scalar, since photon trajectories are also affected by the Hubble parameter ($H$), the distance measurements do get modified by the non-metricity. In figure \ref{fig-plots}, we plot the expansion history and the angular distance to show the modifications caused by small values of the non-metricity parameter.
	
	The effects of the non-metricity (back-reaction) become more pronounced for large redshifts. For small negative values of the non-metricity parameter, the Hubble parameter is becoming larger than its $\Lambda$CDM value. This could prove useful in the context of Hubble tension \cite{dival} in which the cosmic microwave data suggests a smaller value of the Hubble constant than the local observations. However, it is worth noting that the back-reaction parameter has been previously constrained to have small positive values \cite{clarkson,tharake2}.

	\section{Summary and Discussion} \label{sec-disc}
	
	In macroscopic gravity, the averaging procedure breaks the metricity of geometry, and the cosmological back-reaction can be given in terms of the non-metricity of the space-time. Since the non-metricity results from a change in the geometric structure of the space-time, the effects of back-reaction are not limited to dynamics. In this paper, we analysed the effects of non-metricity on the geometric properties of the averaged universe. The primary goal of this paper was to demonstrate how non-metricity modifies kinematics by deriving the Raychaudhuri and Sachs equations. However, the non-Riemannian nature of space-time raises questions about the trajectories of physical particles. It has been argued that only matter that possesses a `microstructure' (properties like hypercharge and dilation) couples with the non-Riemannian geometric fields and thus will be affected by them \cite{obu}. In the metric-affine gravity literature, such matter has been termed as a `hyperfluid' \cite{obu1,ios1}. Ordinary matter (e.g. perfect fluid or dust used in cosmology) does not possess such microstructure. Hence, despite the averaged geometry being non-Riemannian, we took the paths traced by physical particles to be simply Riemannian geodesics.
	
	
	Having clarified this issue, we presented a $1+3$ decomposition of the averaged space-time geometry within macroscopic gravity. We started with deriving the Raychaudhuri and Sachs equations which govern the expansion of the geodesic flows. Physically, these equations characterise the flow of the cosmic fluid and photon bundles, respectively. Both of these equations were modified due to non-metricity. However, we found that only the Riemannian part of the expansion scalar was related to the fractional rate of change of cross sectional area of the geodesic bundles. Then, we presented geometric properties of the spatial hypersurfaces orthogonal to these geodesics. Using the definition of the spatial covariant derivative, we showed that the spatial hypersurfaces are also non-Riemannian in nature. This became even more apparent in the expressions for the extrinsic curvature. Using the Ricci identity for hypersurfaces, we derived the Gauss-Codazzi equations. As expected, they showed us that non-metricity modifies the Riemann curvature projected onto the hypersurfaces.
	
	Finally, we applied the results derived in this paper to observational cosmology. Using the Sachs optical equation, we derived the angular diameter distance formula. We showed that, despite the existence of a non-metricity, the angular diameter distances \textit{do not} explicitly get modified by it. This was a consequence of the cross sectional area depending only on the Riemannian part of the expansion scalar. Therefore, one needs to be careful when using distances for observational cosmology within macroscopic gravity. Including the averaged Ricci tensor directly in the Sachs equation might result in an incorrect evaluation of the effects of back-reaction on distances in cosmology. However, it is important to note that the non-metricity does modify the distances implicitly through the dynamics (Hubble parameter). We plotted the expansion history and the angular distance formula to demonstrate the deviation from their $\Lambda$CDM counterparts for small values of the back-reaction. We note that a small negative value of the back-reaction parameter could help alleviate the Hubble tension. This warrants further investigations into the averaging problem.
	
	\section*{Acknowledgements}
	We thank Mustapha Ishak for useful comments. AA would also like to thank the anonymous reviewer for one of our previous manuscripts whose comments were crucial in the conception of this paper.
	
	\appendix
	
	\section{Constraints on the Connection Correlation} \label{app-conncons}
	To derive a solution to the averaged field equations \eqref{macroefe}, we need to determine the form of the connection correlation for a given macroscopic geometry. This is done by solving the following algebraic and differential constraints on the connection correlation \cite{zala1,zala2},
	\begin{align}
		{Z^\alpha}_{\beta\gamma}{^\mu}_{\nu\sigma} &=  -{Z^\alpha}_{\beta\sigma}{^\mu}_{\nu\gamma} \label{corr2formprop1}\\
		{Z^\alpha}_{\beta\gamma}{^\mu}_{\nu\sigma} &=  -{Z^\mu}_{\nu\gamma}{^\alpha}_{\beta\sigma} \label{corr2formprop2}	\\	
		{Z^\alpha}_{\beta[\gamma}{^\mu}_{\nu\sigma]} &= 0 \label{corr2formprop3}\\
		{Z^\epsilon}_{\epsilon\gamma}{^\mu}_{\nu\sigma} &= 0 \label{corr2formprop4}\\
		{{{Z^\alpha}_{\beta\mu}}^\gamma}_{\delta\nu}u^\nu &= 0 \label{corr2formprop5}\\
		{Z^\alpha}_{\beta[\gamma}{^\mu}_{\underline{\nu}\sigma||\lambda]} &= 0 \label{corr2formprop6}
	\end{align}
	\begin{equation}\label{corr2formprop7}
		{Z^\epsilon}_{\beta[\gamma}{^\mu}_{\underline{\nu}\sigma}{\bar{R}^\alpha}_{\underline{\epsilon}\lambda\rho]} - {Z^\alpha}_{\epsilon[\gamma}{^\mu}_{\underline{\nu}\sigma}{\bar{R}^\epsilon}_{\underline{\beta}\lambda\rho]} + {Z^\alpha}_{\beta[\gamma}{^\epsilon}_{\underline{\nu}\sigma}{\bar{R}^\mu}_{\underline{\epsilon}\lambda\rho]} - {Z^\alpha}_{\beta[\gamma}{^\mu}_{\underline{\epsilon}\sigma}{\bar{R}^\epsilon}_{\underline{\nu}\lambda\rho]} = 0 
	\end{equation}
	where, $\boldsymbol{u}$ is the unit time-like 4-vector field and $\boldsymbol{\bar{R}}$ is the Riemannian part of the curvature tensor for the (assumed) macroscopic metric.   
	
	Equations \eqref{corr2formprop1}-\eqref{corr2formprop5} are algebraic constraints and do not depend on the macroscopic geometry. Solving these leaves one with 121 independent components in the connection correlation. Equations \eqref{corr2formprop5}, \eqref{corr2formprop6} and \eqref{corr2formprop7} ensure that the higher order correlations are zero. One also assumes that the connection correlation is invariant under the same group of isometries as the macroscopic metric \cite{hoogenss}. Given this, the connection correlation can be determined completely. A systematic way of solving the above macroscopic gravity identities has been presented in \cite{hoogenss,hoogen2,tharake1}. The disformation tensor can also be solved for in a similar manner.

	\bibliographystyle{hunsrtnat}	
	\bibliography{averageQ}
	
\end{document}